\documentclass[superscriptaddress,pra]{revtex4}

\usepackage[latin1]{inputenc}
\usepackage{graphicx}
\usepackage{amsmath}
\usepackage{amssymb}
\usepackage{dsfont}
\usepackage{graphicx}

\makeatletter
\def\erf{\mathop{\operator@font erf}\nolimits}
\makeatother

\newcommand\be{\begin{equation}}
\newcommand\ee{\end{equation}}

\begin{document}

\title{Generation of quasi-rectangle-states of the vibrational motion of an ion}
\author{Ir\'an Ramos-Prieto, Iv\'an Rinc\'on,  V\'ictor Arrizon and  H\'ector M. Moya-Cessa}
\affiliation{Instituto Nacional de Astrof\'{i}sica, \'Optica y Electr\'onica, Calle Luis Enrique Erro No. 1, Santa Mar\'{i}a Tonantzintla, Pue., 72840, Mexico}

\begin{abstract}
We show how to generate quasi-rectangle-states of the vibrational motion of an ion, this is, states that have the same probability in a given position interval. We produce those states by repeated ion-laser interactions followed by conditional measurements that generate a superposition of squeezed states. The squeeze parameter of the initial state  may be tuned in order to obtain such highly localized rectangle states. \end{abstract}
 \maketitle

\section{Introduction}
The early papers of Roy Glauber \cite{Glauber1,Glauber2} on coherence, marked a course on how to define nonclassical states, a first good approximation was: Those states that have less fluctuations in a given observable than that of a coherent state \cite{Glauber2}. Since then, a myriad of ways of generating nonclassical states, theoretically and experimentally, have been proposed \cite{Wine,CirZol,Dan,Win1,monroe,Moya1999,kis,mato,vogeli}.

Nonclassical states of the centre-of-mass motion of a trapped ion have played an important role because of fundamental problems in quantum mechanics and for their potential practical applications such as
precision spectroscopy \cite{Wine} and quantum computation \cite{CirZol,Dan}.  By exhibiting less fluctuations than that of a coherent state, the so-called  standard quantum limit they are of great importance. 

Ways of generating squeezed states \cite{Win1}, superpositions of coherent states \cite{monroe,Moya1999},
nonlinear coherent states \cite{kis,mato,vogeli}, number states and some 
specific superpositions of them have been proposed \cite{Moya2000}. In experimental and
theoretical studies of single trapped ions interacting with laser beams it has been usually considered the case in which such interaction may be modeled as a
Jaynes-Cummings interaction \cite{Win1,winl,Blockley}, therefore exhibiting  collapses and revivals \cite{Cir1} and the generation of nonclassical states, peculiar of such a model, or its multiphotonic generalizations \cite{Cir2,Cir3,Blatt}. It should be noted also that, via a similarity transformation that does not need of any approximations, the ion-laser interaction may be taken to  a two-level atom interacting with a quantized field, i.e., the Rabi interaction \cite{Moya2003,Moya2012,Moya2016}. In fact multiphonon Rabi models may be realized via such transformation \cite{Casanova2018}. 

When studying ion-laser interactions,  usually two rotating wave approximations  are performed, the first
related to the laser optical frequency and the second to the vibrational frequency of the ion, in order to remove counter-propagating terms of the Hamiltonian, difficult to  be treated analytically.  Approximations on the Lamb-Dicke parameter, $\eta$, are usually done, considering it much smaller than unity. {In order to perform the rotating wave approximation, the laser intensity, $\Omega$, is considered much smaller than the trapping frequency, $\nu$, namely $\Omega \ll \nu$  \cite{Poyatos}.}

As mentioned above, cavity fields and trapped ions share several common features as in both topics the possibility to realize  Jaynes-Cummings \cite{jaynes-cummings,Shore} and anti-Jaynes-Cummings \cite{Lara2} interactions, are feasible. In cavities interacting with atoms this model describes the resonant or slightly non-resonant  interaction of an atom with the cavity mode \cite{rempe-jcm}.
 On the other hand, for a trapped ion in the Lamb-Dicke regime the Jaynes-Cummings model
describes the  interaction of an electronic transition and the quantized centre-of-mass motion, assisted by a laser beam,  in the resolved sideband
regime~\cite{wineland-jcm}. The ion-laser interaction has an advantage over the atom-field interaction, namely, decoherence processes do not affect the ion-laser interaction as much as it does to atom field interactions where cavities suffer greater losses \cite{Aguilar}.  This means that effects that occur in interacting cavities \cite{Hanoura} may be better produced in ion-laser interactions.

\begin{figure}
    \centering
    \includegraphics[width=.6\linewidth]{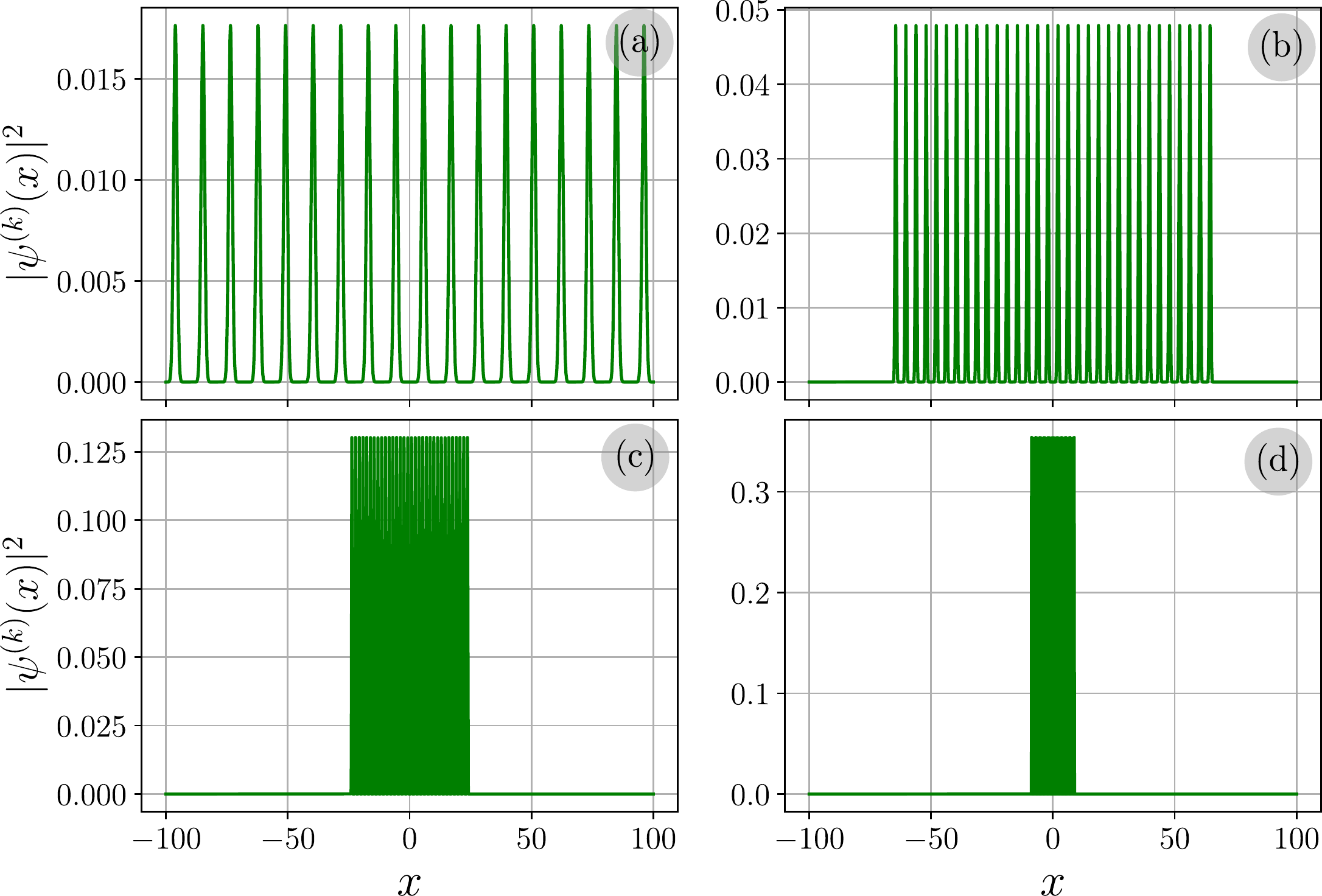}
    \caption{Probability to find the centre-of-mass motion of the ion in a given position for (a) $r=0$, (b) $r=1$, (c) $r=2$ and (d) $r=3$ with  $\tau=4e^{-r}$. The states ploted are obtained after $k=4$ consecutive measurements of the ion in its excited state.  }
\end{figure}
\begin{figure}
    \centering
    \includegraphics[width=.6\linewidth]{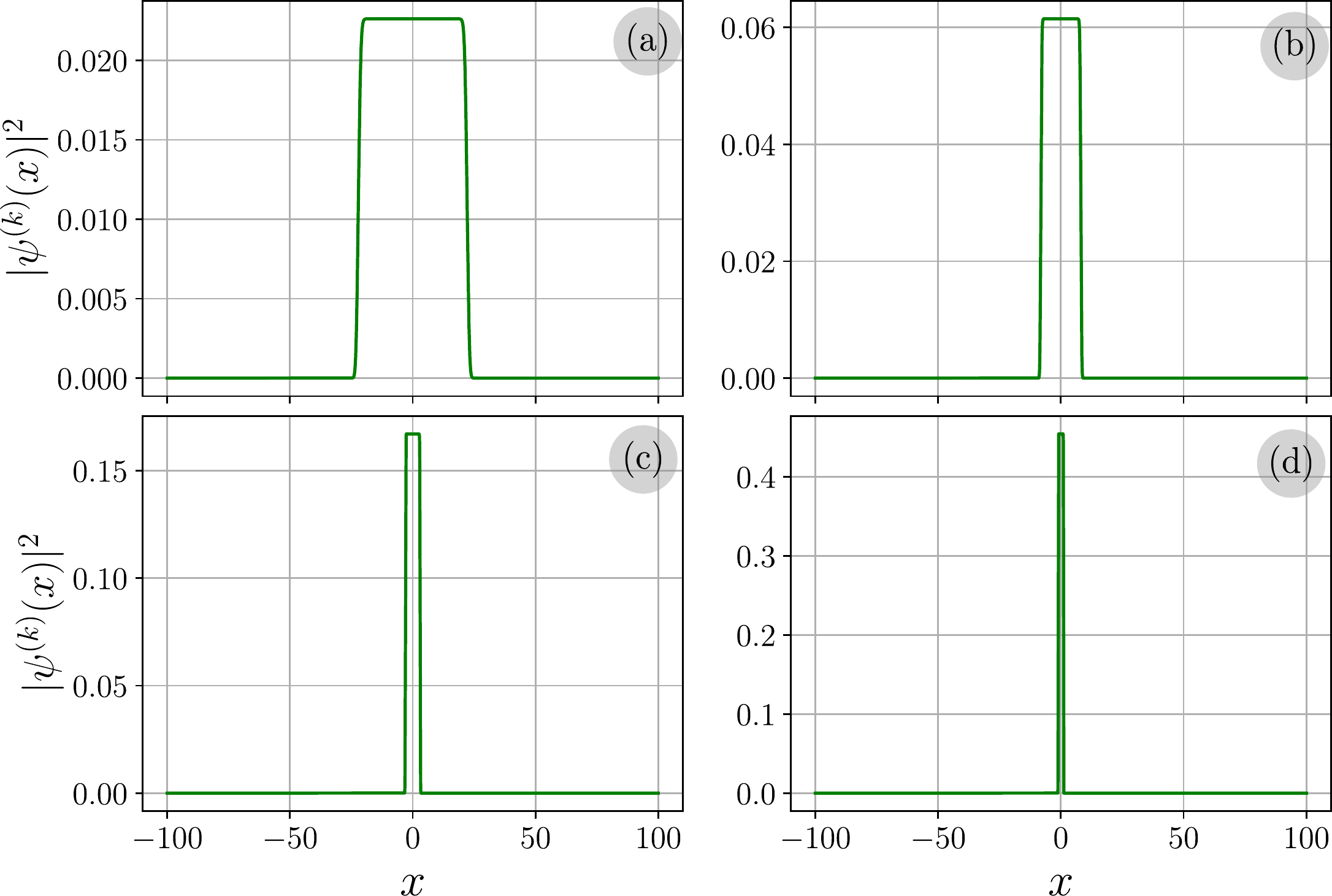}
    \caption{ Probability to find the centre-of-mass motion of the ion in a given position for (a) $r=0$, (b) $r=1$, (c) $r=2$ and (d) $r=3$ with  $\tau=e^{-r}/2$. The states ploted are obtained after $k=4$ consecutive measurements of the ion in its excited state.}
\end{figure}
In this contribution we will show how to generate quasi-rectangle states, this is, states of equal probability in a position interval by a set of conditional measurements \cite{kurizki,Schleich}.
We will consider an initial squeezed vacuum state and, by changing the squeezing parameter, we will be able to produce some non-classical states that have the same probability in configuration space in a given window of positions, i.e., a quasi-rectangle function.

\section{Superpositions of equidistant squeezed states}
{We consider an interaction  proposed previously in the reconstruction of the quantum mechanical state of a trapped ion \cite{Wallen1} and that has been used in the preparation of coherent superpositions of arbitrary quantum states such as Schr\"odinger cats \cite{Wallen2}. A weak electronic transition of the ion is  irradiated by two laser fields detuned to the first lower and first upper vibrational sidebands of the transition, respectively. For equal laser intensities it has been shown, that, for an ion trapped in the resolved-sideband and Lamb-Dicke regimes and in the interaction picture the  Hamiltonian is given  by \cite{Moya1999,Wallen1}}
\begin{figure}
    \centering
    \includegraphics[width=.6\linewidth]{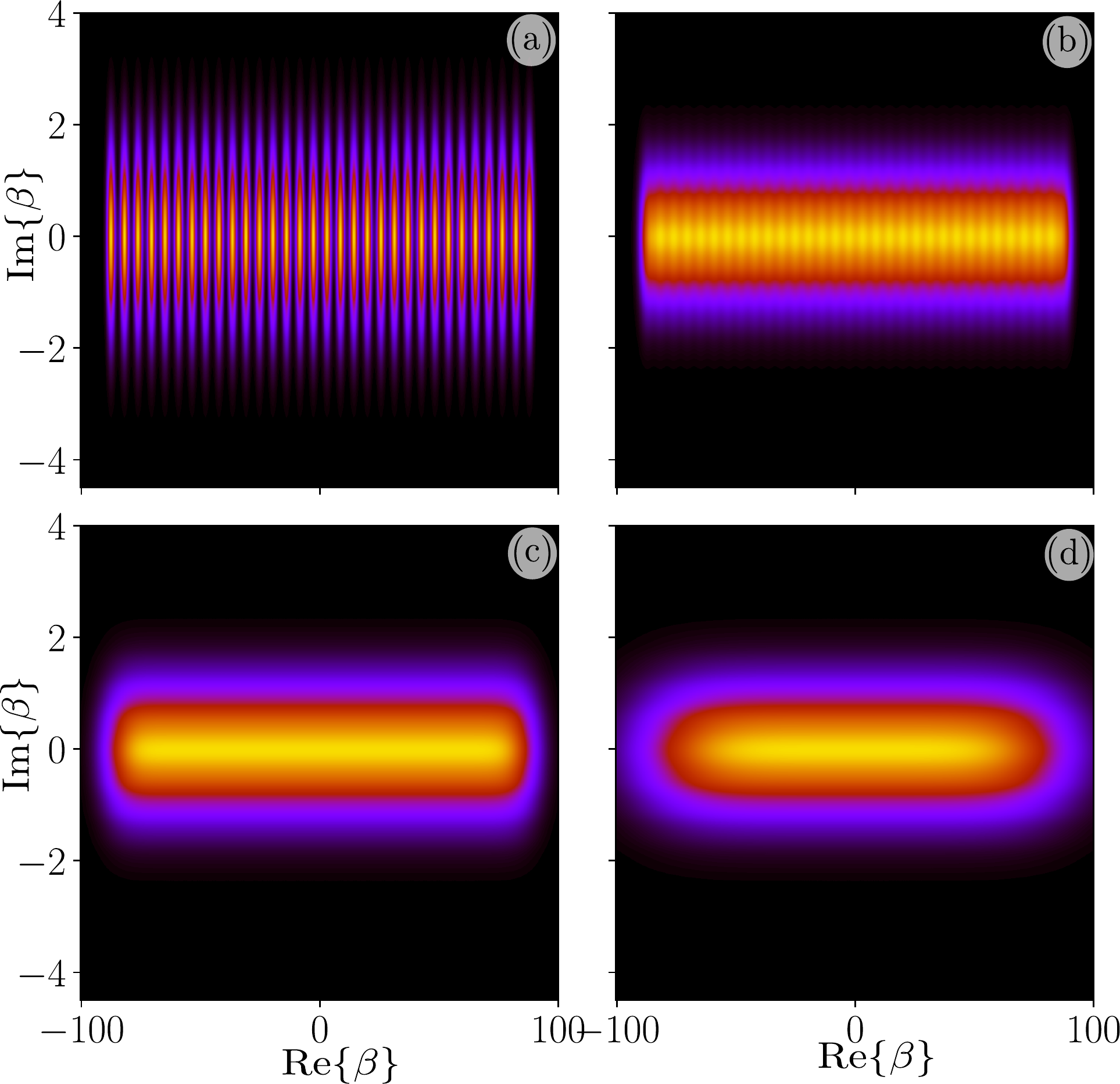}
    \caption{ Plot of the Husimi $Q$-function for the same parameters as in Figure 1.}
\end{figure}

\begin{equation}\label{Ham}
H_I=g(ae^{i\phi}+a^{\dagger}e^{-i\phi})(A_{eg}+A_{ge}).
\end{equation}
{where $A_{eg}$ ($A_{ge}$) is the operator that describes the electronic transition from the ground (excited) state to the excited (ground) state. The effective Rabi frequency is given by the parameter $g$. The annihilation (creation) operator for  the centre-of-mass position of the trapped atom is given by $a$ ($a^{\dagger}$), 
and the phase $\phi$ is determined by the phase difference of the red and blue detuned laser fields. The Hamiltonian (\ref{Ham}) is valid only in the Lamb -Dicke regime, i.e. when the ionic centre-of-mass motion is strongly localized with respect to the laser wavelengths, which is not a serious restriction, as  the Lamb-Dicke parameter may be varied over a wide range by simply varying the geometry of the laser beams that are used to drive the electronic transition in a Raman configuration \cite{Win1,monroe}.}

We set $\phi=\pi/2$ and find easily the evolution operator, $U_I(t)=\exp(-iH_It)$, as
\begin{equation}
    U_I=\cos[igt(a-a^{\dagger})](A_{ee}+A_{gg})-i\sin[igt(a-a^{\dagger})](A_{eg}+A_{ge}).
\end{equation}
By considering the initial state of the whole wave function to be
\begin{equation}
   |\psi(0)\rangle= |0,r\rangle|e\rangle,
\end{equation}
i.e., the vibrational motion given by the squeezed  vacuum state  \cite{Loudon,Yuen,Caves,Nieto,Barnett,Vidiella} 
\begin{equation}
|0,r\rangle=S(r)|0\rangle,
\end{equation}
with $S(r)=e^{-\frac{r}{2}(a^2-a^{\dagger 2})}$, and the internal state of the ion in the excited state, we obtain the evolved wave function as
\begin{equation}
   |\psi(t)\rangle= \cos[gt(a+a^{\dagger})]|0,r\rangle|e\rangle-i\sin[gt(a+a^{\dagger})]|0,r\rangle|g\rangle. \label{first}
\end{equation}

The application of the squeezed operator to the annihilation and creation operators gives
\begin{equation}
S(r)aS^{\dagger}(r)= \mu a + \nu a^{\dagger}, \qquad
S(r)a^{\dagger} S^{\dagger}(r)= \mu a^{\dagger} + \nu a,
\end{equation}
with $\mu=\cosh r$ and $\nu=\sinh r$.
{Because spontaneously emitted photons disturb the motional quantum state via recoil effects, we consider conditional measurements  such that only the excited state is observed. Then the quantum state after measuring the state $|e\rangle$ at a time $t_1$ is given by}
\begin{equation}
   |\psi^{(1)}(t_1)\rangle=\frac{1}{N_1} \cos[igt_1(a-a^{\dagger})]|0,r\rangle|e\rangle,
\end{equation}
where $N_1$ is a normalization constant.

If we consider this state as initial state for a second interaction, we end up with the following wavefunction

\begin{equation}
   |\psi^{(2)}(t_2)\rangle=\frac{1}{N_2} \cos[igt_2(a-a^{\dagger})]\cos[igt_1(a-a^{\dagger})]|0,r\rangle|e\rangle,
\end{equation}
and after  $k$ interactions we obtain the vibrational state
\begin{equation}
|\psi^{(k)}(t_k)\rangle=\frac{1}{N_k} \prod_{j=0}^k \cos[igt_j(a-a^{\dagger})]|0,r\rangle|e\rangle.
\end{equation}
If we choose $gt_j= 2^j\tau$, with $\tau$ a (dimensionless) scaled time,  that we will fix later, we can prove that the vibrational wavefunction reads
\begin{equation}
|\psi^{(k)}(t_k)\rangle_v=\frac{1}{\tilde{N_k}} \sum_{j=0}^{2^k-1} \cos[i\tau (2j+1)(a-a^{\dagger})]|0,r\rangle, 
\end{equation}
with $\tilde{N}_k$ a new normalization constant, related to the number of interactions, $k$, that we may find from the condition 
\begin{equation}
    _v\langle \psi^{(k)}(t_k)|\psi^{(k)}(t_k)\rangle_v=1.
\end{equation}
We can rewrite the above equation as
\begin{equation}
|\psi^{(k)}(t_k)\rangle_v=\frac{1}{\tilde{N_k}}\left( \sum_{j=0}^{2^k-1} |(2j+1)\tau,r\rangle+\sum_{j=0}^{2^k-1} |-(2j+1)\tau,r\rangle \right) \label{vib}
\end{equation}
with $|\alpha,r\rangle$ a squeezed state of amplitude $\alpha$
\begin{equation}
|\alpha,r\rangle=D(\alpha)S(r)|0\rangle, \label{sque}
\end{equation}
with $D(\alpha)$ the Glauber displacement operator \cite{Glauber2}.

The normalization constant is then given by
\begin{equation}
\tilde{N_k}^2=2\sum_{j=0,m=0}^{2^k-1}e^{-{2(j+m+1)^2}\tau^2e^{2r}}+e^{-{2(j-m)^2}\tau^2e^{2r}},
\end{equation}
from the fact that, for  $\alpha_1$ and $\alpha_2$ real, $\langle \alpha_1,r|\alpha_2,r\rangle=e^{-\frac{(\alpha_1-\alpha_2)^2e^{2r}}{2}}$.
\begin{figure}
    \centering
    \includegraphics[width=.6\linewidth]{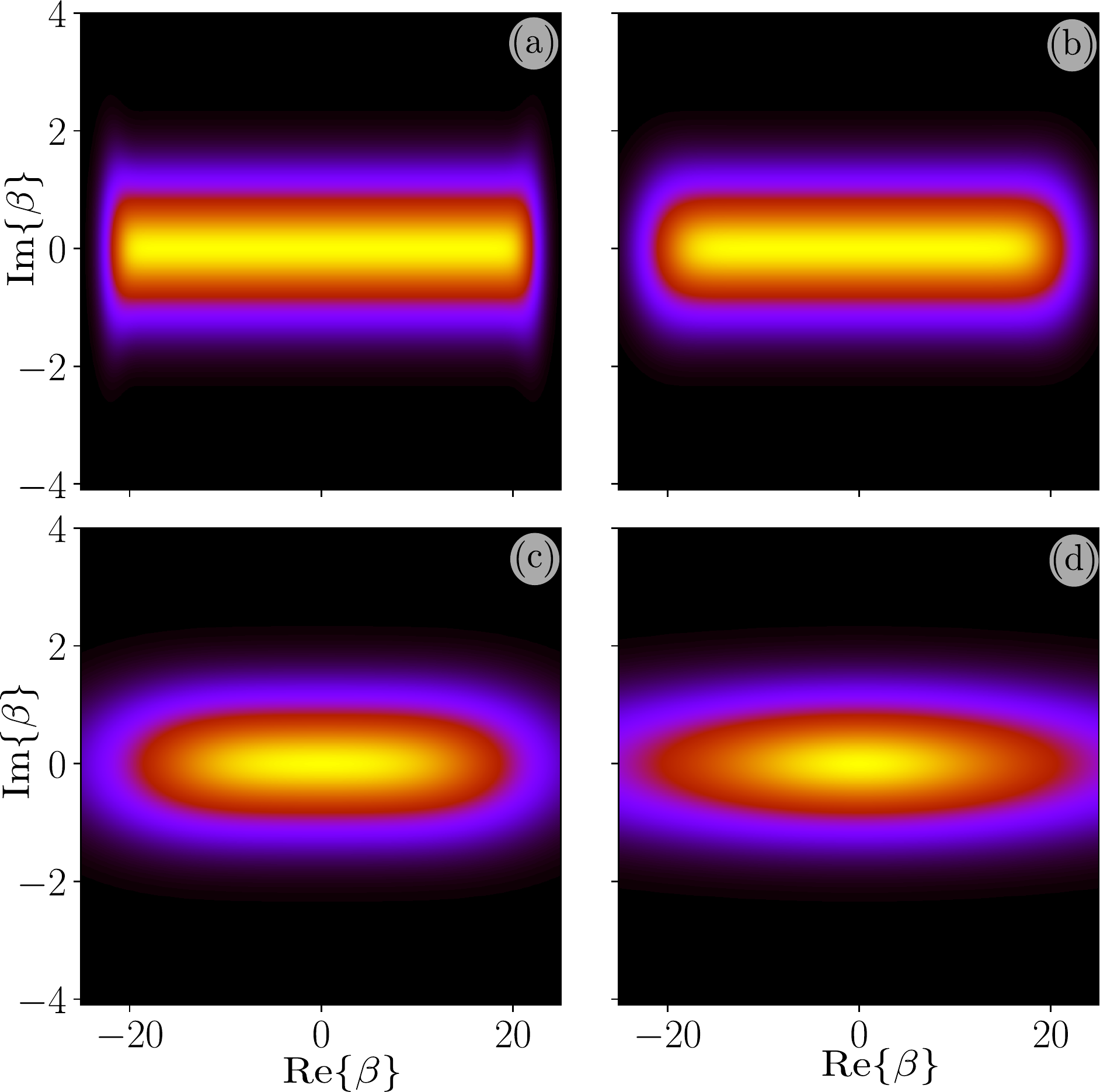}
    \caption{ Plot of the Husimi $Q$-function for the same parameters as in Figure  2.} 
\end{figure}
\section{Quasi-rectangle states}
We now show how, by tuning the squeezing parameter, $r$, a quasi-rectangle state, namely, a state with equal probability in a position window, may be generated. In order to do this, we write the vibrational wavefunction (\ref{vib}) in configuration space
\begin{eqnarray}
\psi^{(k)}(x)&=&_v\langle x|\psi^{(k)}(t_k)\rangle_v\\ &=&\left(\frac{e^{2r}}{\pi}\right)^{1/4}\frac{1}{\tilde{N_k}} \sum_{j=0}^{2^k-1} e^{-\frac{e^{2r}}{2}[x-\sqrt{2}(2j+1)\tau]^2}+e^{-\frac{e^{2r}}{2}[x+\sqrt{2}(2j+1)\tau]^2}\nonumber
\end{eqnarray}
where we have used that, for $\alpha_1$ real
\begin{eqnarray}
\psi(x)=\langle x|\alpha_1,r\rangle=\left(
\frac{e^{2r}}{\pi}\right)^{1/4}e^{-\frac{e^{2r}}{2}(2\alpha_1^2-2\sqrt{2}x\alpha_1+x^2)},
\end{eqnarray}
and $|x\rangle$ is a position eigenstate. In Figs. 1 and 2 we plot the probabilities in configuration space after the ion has interacted four consecutive times and measured in its excited state. In Fig. 1, we use $\tau=4e^{-r}$ for 4 different values of the squezing parameter. It may be seen that Although a reduction of the window where the centre-of-mass motion may be found, the oscillations  due to the nature of the Gaussian states that form the superposition, do not dissapear. Fig. 2 shows that by reducing the parameter $\tau$, the oscillations may be eliminated, generating quasi-rectangle states that may be well localized.

\subsection{Husimi function}
Quasiprobability distribution functions \cite{Glauber2,Wigner,Husimi,Glauber3} are useful in several topics in quantum mechanics. They have allowed the reconstruction of the quantum state of the vibrational motion of an ion \cite{Leibfried} and the quantum state of light in CQED experiments \cite{Bertet}. Besides, they may be used to visualize the states in phase to hint about the possible non-classicality of a state.

The Husimi $Q$-function is one of the simplest to use because, besides being always positive, it has a simple expression
\begin{equation}
Q(\beta)=\frac{1}{\pi} \langle \beta|\rho |\beta\rangle,
\end{equation}
with  $|\beta\rangle$ is a coherent state \cite{Glauber2} and $\rho$ ($=|\psi\rangle\langle\psi|$ for a pure state)  is the so-called density matrix of the system.

For the state (\ref{vib}) its Husimi-$Q$ \cite{Husimi} is given by 
\begin{equation}
Q(\beta)=\frac{1}{\pi\tilde{N_k}^2}\left\vert  \sum_{j=0}^{2^k-1} \langle\beta|(2j+1)\tau,r\rangle+\langle\beta|-(2j+1)\tau,r\rangle \right\vert^2 \label{Qvib},
\end{equation}
with 
\begin{equation}
 \langle\beta|(2j+1)\tau,r\rangle=e^{-\frac{\tau^2(2j+1)^2}{2}-\frac{|\beta|^2}{2}-\frac{\nu\tau^2(2j+1)^2}{2\mu}+\frac{\nu\beta^2}{2\mu}+\frac{\beta\tau(2j+1)}{\mu}}.
\end{equation}
We plot these function in figures 3 and 4 for the states generated in Figures 1 and 2, showing a similar behaviour, namely the reduction of the positions where the centre-of-mass motion of the ion may be found. Note that Fig. 4 has a different scaling than Fig. 3.
\section{Conclusions}
We have shown that non-classical states of the vibrational motion of the centre-of-mass motion of an ion may be generated to produce equal probabilities to find it in a given window of positions. By starting from a squeezed vacuum state, we have shown how to produce a equidistant superposition of squeezed states with equal coefficients such that, by tuning the squeezing parameter, a quasi-rectangle function for the obtained probability in configuration space  may be highly localized.

\end{document}